\documentclass{article}
\usepackage[margin=2.5cm]{geometry}

\usepackage{amsmath,amssymb,amsfonts}
\usepackage{orcidlink}

\usepackage{graphicx,color,epsfig}
\usepackage[nocompress]{cite}

\usepackage{authblk}

\def\be{\begin{equation}}
\def\ee{\end{equation}}
\def\beq{\begin{eqnarray}}
\def\eeq{\end{eqnarray}}
\def\n{\nonumber}

\usepackage{enumitem}

\begin{document}

\renewcommand{\arraystretch}{1.3}
\setlength{\tabcolsep}{1ex}

\title{Dynamical systems approach to stellar modelling in  $f(G, B)$ gravity}

\author[1]{
Sudan Hansraj
\orcidlink{0000-0002-8305-7015}
\footnote{Email: hansrajs@ukzn.ac.za}
}

\author[1,2]{Christian G. B\"ohmer
\orcidlink{0000-0002-9066-5967}
\footnote{Email: c.boehmer@ucl.ac.uk}
}

\author[1]{Ndumiso Buthelezi\orcidlink{}\footnote{Email: 215033901@stu.ukzn.ac.za}}

\affil[1]{
Astrophysics Research Centre, School of Mathematics, Statistics \authorcr and Computer Science, University of KwaZulu-Natal, 
\authorcr Private Bag X54001, Durban 4000, South Africa
\protect\vspace{1ex}
}

\affil[2]{Department of Mathematics, University College London, \protect\\ Gower Street, London WC1E 6BT, United Kingdom
}

\date{23 April 2026}
\maketitle

\begin{abstract}

The novel proposal to invoke the split of the Ricci scalar into bulk and boundary terms in the gravitational action, opens up a new avenue of investigation into stellar dynamics. The Lagrangian contains functional forms of the bulk term while the boundary term do not contribute to the dynamics.  The advantage of the proposition is that the stellar structure equations are  up to order two, thus the theory is not haunted by ghosts. We obtain explicitly the defining equations for the thermodynamical variables and the geometry for  the pure quadratic case, since the linear case amounts to general relativity. In trying to establish the vacuum geometry associated with the theory it turns out that two possible metrics emerge through the vanishing of the energy-momentum tensor. Next, we analyse the isotropy equation and make the observation that it is autonomous. It is rare that this happens in astrophysical modelling. This behaviour prompted the use of dynamical systems to understand the stability properties of fixed points or invariant submanifolds. It was necessary to choose a gauge in order to split the autonomous equation into a system from which we could plot a phase portrait and deduce the stability of solution trajectories. We find that the invariant submanifolds were generally stable with nearby paths approaching them.

\end{abstract}

\section{Introduction}
\label{intro:Sec}

Extensions and modifications of the standard theory of the gravitational field, namely Einstein's general relativity, has generated increased interest in recent times. These are motivated on the grounds have that the standard theory fails to address important problems like explaining the late-time accelerated cosmic expansion without the need for exotic matter. Additionally the theory has hardly been tested at the length scales of the universe and in the vicinity of extreme gravity regimes. Moreover, general relativity is known to be not renormalizable in contrast with the situation in quantum field theory~\cite{Goroff1985,tHooft1974}.  

A wide range of modifications have emerged lately. Amongst the leading ideas is that of Lovelock~\cite{love1,love2} gravity that contains the most important ingredients of a viable theory namely diffeomorphism invariance, compliance with the Bianchi identities, up to second order derivatives in the equations on motion and Ostrogradsky stability. The second order Lovelock polynomial gives rise to Einstein-Gauss-Bonnet (EGB) gravity which finds substantial support from the fact that it appears in the low energy effective action of string theory which itself is a leading quantum field theory. The possible drawback of Lovelock gravity is that higher curvature effects only become dynamic in more than 4 dimensions. Recent efforts to rescale the coupling to accommodate 4 dimensions in EGB gravity has stirred much controversy~\cite{Glavan} although it is known that the proposal is safe in highly symmetric spacetimes such as spherical symmetry~\cite{GursesSismaniTekin}. For the purposes of this work the reader should note that extensions of Gauss-Bonnet gravity includes functional forms of the Gauss-Bonnet invariant $\mathcal{G}$ and goes by the name $f(\mathcal{G})$ gravity. We will distinguish our work by referring to the B\"ohmer-Jensko model~\cite{BJ} as $f(G, B)$ theory even though the boundary term $B$ will not feature. 

Other notable alternatives include $f(R)$ theory which suffers from being haunted by ghosts of the theory in the form of fourth order derivatives~\cite{Nojiri2007,Sotiriou2010,DeFelice2010}. The three theories $f(R, T)$~\cite{Harko2011}, Rastall gravity~\cite{Rastall1972} and unimodular theory~\cite{HansrajEllis1} are basically equivalent to general relativity in a geometric sense and have the disadvantage of displaying non-conservation of energy momentum. The recent concept of non-metricity in the form of $f(Q)$ theory shows little promise as it is only consistent for the linear case and consequently it gives identical results to general relativity.  Most of these ideas have proved successful in explaining the universe's accelerated expansion, however, they include their own weaknesses. Most notably amongst higher than linear orders of $f(R)$ and $f(\mathcal{G})$ gravity is the inability to yield suitable astrophysical models even of the most elementary perfect fluid type. Compact star models are possible, however to facilitate such requires identifying higher order derivative terms originating from the geometric structure with the matter content thereby giving unfamiliar fluid distributions. Physically reasonable models with an equation of state are extremely difficult to obtain. 

In~\cite{BJ} a proposal was put forth to unify classes of modified gravity theories by considering the impact of the boundary terms in the decomposition of the Ricci scalar. They noted that metric-affine theories admit various other formulations  such as teleparallel equivalent of general relativity (TEGR)~\cite{Maluf2013} in which case the action of the theory is Lorentz invariant up to a boundary term. This is also the case for the Einstein-Hilbert action of general relativity. The highly studied $f(R)$~\cite{Starobinsky} theory is invariant under Lorentz and diffeomorphism transformations but has the drawback of generating fourth order differential equations as remarked earlier. B\"ohmer and Jensko~\cite{BJ} proposed that if the Ricci scalar were to be separated into bulk and boundary terms then a theory yielding up to second order derivatives would emerge by considering only the bulk terms in the action. 

Examining boundary terms is not a novel consideration. These were investigated by in~\cite{BJ} where $f(T, B)$ gravity was studied with $T$ being the torsion scalar and $B$ the boundary term. In this context $f(R)$ is found to be the unique Lorentz invariant theory and $f(T)$ gravity as the unique second order theory. The central idea in the work of Boehmer and Jensko is to identify three types of boundary terms which after careful consideration will admit $f(T)$, $f(Q)$ and $f(R)$ gravity theories as limiting cases. An extension of these ideas  by the same authors of a unified approach to modified gravity within the Palatini formalism showing the role of torsion or non-metricity for nonlinear boundary terms may be found in~\cite{BJ2}. 

As a result of the seminal work~\cite{BJ}, several connections between modified theories and boundary terms were discussed. Capozziello {\it{et al}}~\cite{Capozziello2023} investigated the role of the boundary term in $f(Q, B)$ theory and contrasting with $f(Q)$ gravity in the limit $B \to 0$. Shabani {\it{et al}}~\cite{Shabani2024} studied the cosmological significance of the boundary term in symmetric teleparallel gravity using a dynamical systems approach. Similarly Kadam {\it{et al}}~\cite{Kadam2023} analysed extended  teleparallel gravity with a boundary term using a phase plane analysis. In the same spirit Vishwakarma and Shah~\cite{Vishwakarma2024} considered higher curvature gravity models with boundary terms with a log-square-root and a hyperbolic tangent power model.  

Isotropic stars are stellar models of stars where the pressure is the same in all directions inside the star radially and tangentially. This means that the pressure acting outward from the centre is equal to the pressure acting in the tangential direction. Because of this assumption, the matter inside the star can be considered as a perfect fluid. This makes the equations of general relativity harder to solve as a constraint equation arises. However, many exact solutions have been obtained for the spherically symmetric case, see Delgaty and Finch-Skea~\cite{finchskea}. For this reason, isotropic stellar models are very important in theoretical studies and are often used as a starting point when studying compact stars such as neutron stars.

Although real stars may not be perfectly isotropic at very high densities, isotropic models remain useful because they help us understand the basic physical properties of relativistic stars and provide a reference for more complex models. Many researchers have built isotropic stellar solutions in general relativity and in modified gravity theories, including Gauss--Bonnet inspired models~\cite{Tolman1939, Buchdahl1959}. Dadhich {\it {et al}}~\cite{N2Dadhich}, Hansraj {\it {et al}}~\cite{Hansrajt1} and Ivanov~\cite{Ivanov2002} show that isotropic stars are crucial for the understanding compact objects.

The dynamical systems method is a useful way to study differential equations when finding exact solutions is difficult or impossible. The basic idea is to rewrite a complicated differential equation as an autonomous system. Once the system is in autonomous form, we can separate it into two first order equations in new variables. The next step is to find the fixed points or fixed curves of the system, which are locations in the phase space where the derivatives vanish. These fixed points represent exact solutions in the original variables. Because the system is nonlinear in our case, the stability of these fixed points is not immediately obvious.  The phase portrait allows us to visualise the behaviour of solutions without needing explicit formulas~\cite{BanasiakArlotti}. Trajectories in the phase plane show how quantities evolve as the independent variables changes, revealing attractors, repellers, and neutral directions. This approach provides a global picture of the solution space and helps us understand important features of the stellar models, such as the relation between the metric potentials and the overall structure of the spacetime, even when exact solutions are hard to obtain~\cite{Nilsson2001,Heinzle2003}.

There are distinctly 3 main benefits of the approach we follow.  First, to the best of our knowledge this is the first application of the Boehmer--Jensko $f(\mathcal{G},B)$ framework to the modelling of \emph{stellar} interiors; existing studies within this theory have focussed exclusively on cosmological settings. Second, the reduction of the stellar pressure-isotropy condition to an autonomous ordinary differential equation --- achieved here through the scale-invariant substitution $U = r\mu'$, $V = r\nu'$ --- is a rare occurrence in relativistic astrophysics. In standard general relativity the analogous isotropy equation does not admit an autonomous form in these variables, so the present theory offers a qualitatively richer solution structure. Third, and most importantly from a physical standpoint, the dynamical systems analysis reveals that the fixed curves are generically transversely attracting: nearby solution trajectories in the $(\mu,\nu)$-space are drawn toward the self-similar profiles $\mu'(r)\sim r^{-1}$, $\nu'(r)\sim r^{-1}$. This attractor property implies that physically reasonable metric potentials --- even those not belonging exactly to the fixed-curve family --- will inherit the asymptotic behaviour dictated by the invariant submanifolds, thereby constraining the class of viable stellar geometries without requiring exact solutions. The scale-invariant property does indeed suggest a search for exact solutions where the metric potentials vary inversely with the radius. This is the subject of a further investigation.

This work is organised as follows: We introduce the theory of $f(G,B)$ gravity and derive the field equations of the theory for a spherically symmetric background spacetime geometry. Next the pure quadratic model is studied by ignoring the Einstein terms and focusing on the quadratic term only as the next level of complexity. Imposing pressure isotropy, the master field equation is derived. It turns out that  two vacuum solutions are possible, one with flat spatial slices and the other with curved spacetime but containing a curvature singularity, thus illustrating the richer vacuum results of quadratic $f(G, B)$ theory compared to general relativity. Finally, isotropic stellar models in quadratic $f(G,B)$ form are studied using a dynamical systems approach. The pressure isotropy is reformulated in terms of scale-invariant variables, leading to an autonomous system. Fixed points and fixed curves are identified and studied using phase plane analysis. This approach gives an insight into the behaviour of the metric functions without requiring exact solutions. While the split of the master equations is not unique the dynamical systems approach conveys useful insights about the solutions and their stability properties.

\section{The \texorpdfstring{$f(G, B)$}{f(G, B)} gravity formalism}

A novel approach to modified gravity was proposed in~\cite{BJ}, who revisited the Einstein-Hilbert action of General Relativity with a focus on boundary contributions. In their formulation, the Ricci scalar $R$,  can be split into a bulk term $G$ and a boundary term $B$, leading to a second-order theory which can avoid the introduction of ghosts in the form of higher derivative terms. This split is not novel and has been understood since the early days of general relativity. However, it did not receive detailed attention certainly not in the area of astrophysics.  The standard Einstein-Hilbert action reads
\begin{align}
  S_{\rm EH}[g_{\mu \nu}] = \frac{1}{2 \kappa} \int R \sqrt{-g}\, d^4x \,.
\end{align}
Separating the bulk and boundary, the Ricci scalar or curvature scalar $R$, may be  expressed in the fform 
\[
  R = G + B 
  \]
in terms of the bulk $G$ and boundary $B$. Consequently, the Einstein-Hilbert action can be rewritten in terms of the bulk and boundary contributions as
\begin{align}
  S_{\rm EH}[g_{\mu \nu}] =\frac{1}{2 \kappa}  \int R \sqrt{-g} \, d^4x =
  \frac{1}{2 \kappa}  \int \bigl( G + B \bigr) \sqrt{-g} \, d^4x \,.
\end{align}
The bulk term $G$ is quadratic in the connection coefficients or Christoffel symbols and may be expressed as
\begin{align}
    G &= g^{\mu \nu}\Big( \Gamma^{\lambda}_{\mu \sigma} \Gamma^{\sigma}_{\lambda \nu} -
  \Gamma^{\sigma}_{\mu \nu} \Gamma^{\lambda}_{\lambda \sigma} \Big) \,
\end{align}
where the quadratic behaviour is seen within the brackets amongst the connection coefficients $\Gamma$. The boundary term $B$ is second order in the metric derivatives and has the form 
\begin{align}
    B =  \frac{1}{\sqrt{-g}} \partial_{\nu}
  \Bigl( \frac{\partial_{\mu}(g g^{\mu \nu})}{\sqrt{-g}} \Bigr) =
  \frac{1}{\sqrt{-g}} \partial_{\sigma} (\sqrt{-g} B^{\sigma}) = \nabla_{\sigma} B^{\sigma} \,,
\end{align}
where we introduce the notation 
\begin{align}
    B^{\sigma} = g^{\mu \nu} \Gamma^{\sigma}_{\mu \nu} - g^{\sigma \nu} \Gamma^{\lambda}_{\lambda \nu} \,.
\end{align}
The boundary terms may be ignored for stellar modelling as they do not contribute to the dynamics however we mention them here for completeness. This split makes it possible to build more general gravitational theories of the form $f(G, B)$. These theories give more flexibility to describe cosmological and astrophysical phenomena while keeping the field equations second-order, which can avoid unwanted ghost modes. The boundary term $B$ naturally accounts for effects at the edges of spacetime, which are usually ignored in standard gravity. This idea is also inspired by theories that include torsion and non-metricity, where geometry beyond curvature plays an important role in describing gravity.

\section{Field equations of \texorpdfstring{$f(G, B)$}{f(G, B)} gravity}
\label{sec3}

We commence with a Cartesian static and spherically symmetric metric in isotropic form 
\be
ds^2 = -e^{\nu}dt^2 + e^{\mu }\left(dx^2 + dy^2 +  dz^2\right) 
\label{1}
\ee
where $\nu = \nu (r)$ and $\mu = \mu (r)$ are the metric potentials to be determined and $r = \sqrt{x^2 + y^2 + z^2}$. These potentials will later be determined from the field equations. 

Consider the static, spherically symmetric line element written in Cartesian–isotropic form in~(\ref{1}). We may define standard spherical coordinates by 
\begin{equation}
\label{eq:cart-to-sph}
x=r\sin\theta\cos\phi,\qquad
y=r\sin\theta\sin\phi,\qquad
z=r\cos\theta,
\end{equation}
subject to \[ \qquad r>0,\;\,\,\, \theta\in(0,\pi),\; \, \, \, \phi\in(0,2\pi) \]
whence it  is now straightforward to confirm the identity
\begin{equation}
\label{eq:flat-identity}
dx^{2}+dy^{2}+dz^{2} \;=\; dr^{2} + r^{2}\,d\theta^{2} + r^{2}\sin^{2}\theta\,d\phi^{2}
\end{equation}
Substituting \eqref{eq:flat-identity} into \ref{1} yields the spherical–isotropic form
\begin{equation}
\label{eq:metric-sph}
ds^{2} \;=\; -\,e^{\nu(r)}\,dt^{2} \;+\; e^{\mu(r)}\Big(dr^{2} + r^{2}d\theta^{2} + r^{2}\sin^{2}\theta\,d\phi^{2}\Big).
\end{equation}
Equations~(\ref{1}) and~\eqref{eq:metric-sph} are therefore related by a coordinate transformation on the spatial manifold; they represent the same spacetime metric.

The quantity  $G$  evaluates to 
\be 
G =\frac{1}{2} e^{-\mu } \mu ' \left(\mu '+2 \nu '\right) 
\label{G} 
\ee
for the  metric~(\ref{1}). The prime denotes differentiation with respect to $r$.

The differential equations governing the behaviour of the gravitational field were established in the work of Boehmer and Jensko~\cite{BJ} through variation of the action against the metric tensor. These equations may be written out explicitly as
\beq
-\frac{1}{2}e^{-2 \mu } f_{GG}\mu'\left(2 \mu ' \nu ''-2\mu '^2 \nu '-  \mu '^3+  2\mu '' \nu '+ 2\mu ' \mu ''\right) \n \\ \n \\ -\frac{1}{2}e^{-\mu } f_G \left( 2\mu ''+  \mu ' \nu '+  \mu '^2+\frac{4   \mu '}{r}\right) +\frac{f}{2}&=&\kappa  \rho  \label{101a}  \\ \n \\
\frac{e^{-\mu} }{2r} f_G \left( 2\mu' + r\mu'^2 + 2\nu' + 2r\nu' \mu'   \right)-\frac{ f}{2}&=&\kappa  p_r  \label{101b} \\ \n \\
 -\frac{1}{4  } \left(e^{-2 \mu } \left( f_{GG} \left(\mu '+\nu '\right) \left(-2 \mu '' \nu '+2 \mu '^2 \nu '+\mu '^3-2 \mu ' \left(\mu ''+\nu ''\right)\right) \right. \right. && \n \\ \n \\ \left. \left.
+\frac{e^{-\mu } f_{G}}{4r}  \left(r \left(2 \left(\mu ''+\nu ''\right)+\nu '^2\right)+2 \mu ' \left(r \nu '+1\right)+r \mu '^2+2 \nu '\right)\right)\right) -\frac{f}{2} &=& \kappa  p_t  \label{101c}
\eeq
for any choice of function $f({G})$. The pressure isotropy condition 
\begin{multline}
    r f_{GG} \left(\mu '+\nu '\right) \left(-2 \mu '' \nu '+2 \mu '^2 \nu '+\mu '^3-2 \mu ' \left(\mu ''+\nu ''\right)\right) \\ 
    +e^{\mu} f_G \left(-2 r \left(\mu ''+\nu ''\right)+2 \mu ' \left(r \nu '+1\right)+r\mu '^2-r \nu '^2+2 \nu '\right) = 0 
    \label{101d}
\end{multline}
results from the tangential and radial pressures being equated that is setting~(\ref{101b}) =~(\ref{101c}).

Note that the linear choice $f(G) = G$ recovers the familiar Einstein equations
\beq
 -\frac{1}{2}e^{-\mu }  \left( 2\mu ''+  \mu ' \nu '+  \mu '^2+\frac{4   \mu '}{r}\right) +\frac{G}{2}&=&\kappa  \rho \label{102a}  \\ \n \\
\frac{e^{-\mu} }{2r}  \left( 2\mu' + r\mu'^2 + 2\nu' + 2r\nu' \mu'   \right)-\frac{ G}{2}&=&\kappa  p_r  \label{102b} \\ \n \\
\frac{e^{-\mu } }{4r}  \left(r \left(2 \left(\mu ''+\nu ''\right)+\nu '^2\right)+2 \mu ' \left(r \nu '+1\right)+r \mu '^2+2 \nu '\right) -\frac{G}{2} &=& \kappa  p_t  \label{102c}
\eeq
of general relativity in the present coordinate system. Inserting the expression for ${G}$~(\ref{G}) into~(\ref{102a})--(\ref{102c}) gives 
\beq
- e^{-\mu} \left( \mu'' + \frac{2\mu'}{r} + \frac{\mu'^2}{4} \right) &=& \kappa \rho   \\ \n \\
e^{-\mu} \left( \frac{\mu'}{r} +\frac{\nu'}{r} + \frac{\mu'^2}{4} + \frac{\mu' \nu'}{2} \right) &=& \kappa p_r  \\ \n \\
e^{-\mu} \left( \frac{\mu''}{2} +\frac{\nu''}{2} + \frac{\nu'^2}{4} + \frac{\mu'}{2r} +\frac{\nu'}{2r} \right) &=& \kappa p_t \label{102d}
\eeq
after simplification.
The pressure isotropy assumes the form 
\beq
\frac{(\mu' + \nu')}{r} - \frac{\nu'^2}{2} +\frac{\mu'^2}{2} +\mu'\nu'- (\mu'' + \nu'') = 0 \label{isotropy}
\eeq
when the radial and tangential pressures are equated. 
These are precisely the Einstein field equations in 4D but in our chosen coordinate scheme. The equation of pressure isotropy~(\ref{isotropy}) determines the geometry of the model completely. Generally spherical coordinates are employed in the study of stellaar structure and approximately 120 exact solutions of the isotropy equation have been reported in the literature ~\cite{finchskea,delgaty}. Accordingly any of these solutions found will solve the isotropy equation therefore there would be little interest in finding new exact solutions for this case. To make this more transparent an exact solution of the standard Einstein isotropy equation satisfies~(\ref{101d}) and since the left hand side of the isotropy equation~(\ref{isotropy}) appears in the second term of equation~(\ref{102c}), this term vanishes and hence it follows that $f_{GG} =0$ whence we get $f(G) = vG + w$ for $v, w$ some constants. This is equivalent to the Einstein case anyway.  Notably, equation~(\ref{102a}) reduces to an equation expressing the density $\rho$ only in terms of the potential $\mu$. Setting $\rho = $ constant yields the well known Schwarzschild interior metric while the case $\rho = 0$ results in the vacuum Schwarzschild exterior solution in our coordinates. It is straightforward to show the equivalence with the standard spherical coordinates.

The conservation equation, also known as the equation of hydrodynamical stability,  $T^{ab}{}_{;b} = 0$ results in the relationship 
\be 
    2rp_r' = (2r\mu' +4)p_t - (2r\mu' + r\nu' + 4) p_r - r\nu' \rho
\ee
for anisotropic stresses and for isotropy
\be
p' = -\left(\rho + p\right)\frac{\nu'}{2} 
\ee
when $p_r = p_t$. Observe that this relationship is the same as for standard general relativity. 

It is important to note the logical structure of the solution programme. Once a pair of metric potentials $(\mu, \nu)$ satisfying the isotropy equation~(14) is identified --- either exactly or as an attractor trajectory from the phase-plane analysis --- the thermodynamic variables follow algebraically from the field equations: the energy density $\rho$ is given directly by equation~(36), and the isotropic pressure $p$ by equation~(37). The conservation equation~(23) is then automatically satisfied for any solution of the isotropy condition, since it is a differential consequence of the contracted Bianchi identity. A physically complete stellar model additionally requires the specification of an equation of state $p = p(\rho)$. The present work identifies the admissible solution manifolds in metric-potential space; the imposition of a barotropic equation of state such as the linear form $p = \gamma\rho$ or a polytropic relation $p = K\rho^{1+1/n}$ will further select particular trajectories from those manifolds, and this matching will be pursued in subsequent work. It must be observed though that imposing even the simplest equation of state, namely the linear type, creates a differential relationship between the metric potentials  which must be solved simultaneously with the isotropy equation. This is almost impossible and a comprehensive treatment in a separate project will involve numerical techniques.

\section{The special case \texorpdfstring{$G=0$}{G=0}}

The vanishing of the quantity $G$ introduces some intriguing geometrical possibilities that we now consider at the outset. A similar idea was used in~\cite{Tamanini:2012hg} in the context of $f(T)$ gravity. From~(\ref{G}) we note that there are two branches of solutions when $G = 0$.  For the first branch, we set 
\[
\mu' = -2 \nu', \quad \text{so that } \quad \mu = -2\nu + C_3,
\]
where $C_3$ is an integration constant. The metric structure then assumes the simplified form 
\be
ds^2 = -e^{-\frac{\mu}{2}}dt^2 + K e^{\mu}\left(dx^2 + dy^2 + dz^2\right),
\quad r=\sqrt{x^2+y^2+z^2}
\label{4}
\ee

where $K$ is a constant and a further constant has been absorbed into $dt^2$.  The metric potential $\mu$ influences the temporal and spatial behaviour but differently. Clearly, the spacetime manifold is not conformally flat.  With this simplification the system of field equations reduces to 
\beq
-e^{-\mu } f_G \left( \mu '' +\frac{\mu '^2}{4}  +\frac{2   \mu '}{r}\right) +\frac{f}{2}&=&\kappa  \rho \label{3.1}  \\ \n \\
\frac{e^{-\mu }f_G}{2r} \mu' -\frac{f}{2}&=&\kappa  p_r  \label{3.2} \\ \n \\
\frac{e^{-\mu } f_{G}}{4} \left(\mu '' +  \frac{\mu '^2}{4}  + \frac{\mu '}{r}\right) -\frac{f}{2} &=& \kappa  p_t.  \label{3.3} 
\eeq
The pressure isotropy condition $p_r = p_t$ then leads to the differential equation
\be
2\mu'' + \frac{\mu'^2}{2} - \frac{2\mu'}{r} = 0  \label{3.4}
\ee
with general solution
\be
e^{\mu} = C_2 \left( r^2 + C_1 \right)^4  
\label{3.5}
\ee
where $C_1, C_2>0$ are constants, and we choose the domain $r^2 + C_1 > 0$ to avoid coordinate singularities. The corresponding metric reads 
\be
ds^2 = - \frac{\sqrt{C_2}}{\left( r^2 + C_1 \right)^2 }dt^2 + K\left( r^2 + C_1 \right)^4 \left(dx^2 + dy^2 + dz^2 \right),  \quad r=\sqrt{x^2+y^2+z^2} \label{3.5b}
\ee
for isotropic particle pressures. Note that this metric is independent of the function $f(G)$.

To find the vacuum metric for this peculiar case we set each of $\rho, p_r, p_t$ to vanish. Eliminating $f$ from equations~(\ref{3.1}) and~(\ref{3.2}) gives the condition
\be
\mu'' + \frac{\mu'^2}{4} + \frac{\mu'}{r} = 0
\ee
while the same process using~(\ref{3.2}) and~(\ref{3.3}) results in the differential equation
\be
\mu'' + \frac{\mu'^2}{4} - \frac{\mu'}{r} = 0
\ee
and reconciling these requires $\mu' =0$ and hence $\nu' =0$. Consequently we are back to the Minkowski spacetime and the vacuum metric breaks down. 

The second branch from $G = 0$ gives $\mu' = 0$. Then the vanishing density in the vacuum condition forces $f = 0$. Setting the pressures to zero also demand $\nu' = 0$ hence a flat space is the only possible outcome. This effectively rules out compact closed objects but not cosmological fluids. It may still be possible for such a metric to describe a fluid filled universe but we are interested in stellar configurations in this study. This concludes the examination of the case $G = 0$. No viable closed compact objects can exist within these geometries.

\section{The quadratic form \texorpdfstring{$f(G) = \alpha G + \epsilon G^2$}{f(G) = α G + ϵ G²}} 

Since the linear case of $f$ is trivially Einstein, we probe the next level of complexity that is the quadratic form $f(G) = \alpha G + \epsilon G^2$, where the parameters $\alpha$ and $\epsilon$ are real numbers, not necessarily positive. The field equations may be expressed in the form
\beq
8 \kappa  r\rho &=& e^{-2 \mu } \epsilon  \mu ' \left(-32 r \mu '' \nu '+8 \mu '^2 \left(r \nu '-2\right)+5 r \mu '^3 \right. \n  \\ && \left.  -4 \mu ' \left(6 r \mu ''+4 r \nu ''+r \nu '^2+8 \nu '\right)\right)     -2 \alpha  e^{\mu } \left(4 r \mu ''+\mu ' \left(r \mu '+8\right)\right) \label{106a} \\ \n \\
8 \kappa  r p_r &=&  e^{-2 \mu } \left(2 \alpha  e^{\mu } \left(\mu ' \left(r \mu '+2 r \nu '+4\right)+4 \nu '\right) \right. \n  \\ && \left.  +\epsilon  \mu ' \left(\mu '+2 \nu '\right) \left(\mu ' \left(3 r \mu '+6 r \nu '+8\right)+8 \nu '\right)\right) 
 \label{106b} \\ \n \\
8 \kappa  r p_t &=& e^{-2 \mu } \left(2 \alpha  e^{\mu } \left(2 r \left(\mu ''+\nu ''\right)+2 \mu '+r \nu '^2+2 \nu '\right)   \right. \n \\ && \left. +\epsilon  \left(8 r \mu '' \nu '^2+\mu '^3 \left(4-8 r \nu '\right)-3 r \mu '^4+2 \mu '^2 \left(6 r \left(\mu ''+\nu ''\right)  \right.\right. \right. \n \\ && \left. \left. \left.  -r \nu '^2+6 \nu '\right)  
+4 \mu ' \nu ' \left(6 r \mu ''+4 r \nu ''+r \nu '^2+2 \nu '\right)\right)\right).  
\label{106c}
\eeq
The general quadratic form is very complicated so we desire to isolate strictly the quadratic part for the purposes of this study. The equations~(\ref{106a})--(\ref{106c}) accordingly simplify to the system 
\beq
8 \kappa  r\rho &=& e^{-2 \mu } \epsilon  \mu ' \left(-32 r \mu '' \nu '+8 \mu '^2 \left(r \nu '-2\right)+5 r \mu '^3  \right. \n \\ && \left. 
-4 \mu ' \left(6 r \mu ''+4 r \nu ''+r \nu '^2+8 \nu '\right)\right)   \label{107a} \\ \n \\
8 \kappa  r p_r &=&  e^{-2 \mu } \epsilon  \mu ' \left(\mu '+2 \nu '\right) \left(\mu ' \left(3 r \mu '+6 r \nu '+8\right)+8 \nu '\right) \label{107b} \\ \n \\
8 \kappa  r p_t &=& e^{-2 \mu }   \epsilon  \left(8 r \mu '' \nu '^2+\mu '^3 \left(4-8 r \nu '\right)-3 r \mu '^4+2 \mu '^2 \left(6 r \left(\mu ''+\nu ''\right)-r \nu '^2+6 \nu '\right) \right. \n \\  && \left. +4 \mu ' \nu ' \left(6 r \mu ''+4 r \nu ''+r \nu '^2+2 \nu '\right)\right)  \label{107c}
\eeq
when we set $\alpha = 0$. 
The equation of pressure isotropy $p_r = p_t$ assumes the form 
\begin{multline}
    4 r \mu '' \nu '^2-2 \mu '^3 \left(5 r \nu '+1\right)-3 r \mu '^4 +
    \mu '^2 \left(6 r \left(\mu ''+\nu ''\right) \right. \\ 
    \left. -7 r \nu '^2-6 \nu '\right) +2 \mu ' \nu ' \left(6 r\mu '' +
    4 r \nu ''+r \nu '^2-2 \nu '\right) = 0   
    \label{107d}
\end{multline}
for the simplified pure quadratic case. Equation~(\ref{107d}) is the master equation for our study. It encapsulates the full geometry of the system. Gravitational potentials satisfying~(\ref{107d}) may then be used to generate all aspects of the physical stellar model. 

Before proceeding we comment on the connection between the present quadratic theory and standard general relativity. The general quadratic Lagrangian $f(\mathcal{G}) = \alpha\mathcal{G} + \epsilon\mathcal{G}^2$ contains two sectors. Setting $\epsilon = 0$ and retaining $\alpha \neq 0$ reduces $f(\mathcal{G})$ to the linear case, and the field equations~(33)--(35) collapse to the Einstein equations~(15)--(17) derived in Section~3. Thus general relativity is smoothly recovered in the limit $\epsilon \to 0$. The pure quadratic sector studied below, obtained by setting $\alpha = 0$, therefore represents the \emph{leading-order departure} from Einstein gravity within this class of theories. The parameter $\epsilon$ carries dimensions of $[\text{length}]^2$ and controls the strength of the higher-curvature correction; in the regime $|\epsilon|\mathcal{G} \ll 1$ the quadratic contribution is negligible and the classical stellar structure is recovered, while significant deviations are expected at the high curvatures characteristic of compact objects such as neutron stars.

\section{Vacuum branches}

The first important problem to settle is the vacuum solutions of the pure quadratic form of $f(G, B)$ theory. In general relativity, a vacuum is defined as a region where the stress-energy tensor $T_{ab} = 0$. In that framework this is equivalent to the vanishing of the Ricci tensor however this is not the case in $f(G, B)$ gravity. Setting $R_{ab} = 0$ is necessary but not sufficient to determine the vacuum geometry. The vacuum $T_{ab} = 0$ is obtained by setting $\rho = p = 0$ in equations~(\ref{107a}) and~(\ref{107b}). After eliminating $\epsilon e^{-2\mu}$ these equations read as 
\beq
8\,v^{3}\,(r u-2)+5 r\, v^{4}-32r\,v' v\,u-2 v^{2}\!\left(r(12 v'+8 u')+2 r u^{2}+16 u\right) &=& 0,  
\label{EE1}\\[1ex]
-3 r\, v^{2}(v+2 u)+4\!\left(2r(v'+u')+4v(r u+1)+2 r v^{2}+r u^{2}+4 u\right)\,(2 v u+v^{2}) && \n \\  -8 r\left(-2 v' u+2 v^{2}u+v^{3}-2v(v'+u')\right) &=& 0. 
\label{EE2}
\eeq
in which the highest derivatives of the potentials appear as $v'$ and $u'$ after we defined $\mu' = u$ and $\nu' = v$. Essentially the system is first order in the derivatives. This now enables us to  write~(\ref{EE1}) and~(\ref{EE2}) in the form
\beq
A_1(v,u)\,v' \;+\; B_1(v,u)\,u' \;+\; C_1&=&0 \label{EE3} \\
A_2(v, u)\,v' \;+\; B_2(v, u)\,u' \;+\; C_2&=&0 \label{EE4}
\eeq
where
$
A_1 = -r\,(24 v^{2}+32 v u),
B_1 =-16 r\, v^{2}, 
A_2 =8 r\,(v^{2}+2 v u+2 v+2 u),
B_2 = 8 r\, v\,(v+2 u+2).  
$
and $C_{1},C_{2}$  are expressions not containing   $v',u'$ given by 
\beq
C_1 &=&  r\!\left(5v^{4} + 8v^{3}u - 4v^{2}u^{2}\right) - 16v^{3} - 32v^{2}u  \label{C1}    \\
C_2 &=&  8r v^{4} + 32r v^{3}u - 11r v^{3} - 22r v^{2}u + 36r v^{2}u^{2} + 8r v u^{3} + 16 v^{3} + 48 v^{2}u + 32 v u^{2}.  \label{C2}
\eeq 

Now, along any smooth branch $F(v, u) = 0$, the tangent direction must satisfy $(v', u')$ is proportional to $\left(-\frac{\partial F}{\partial u}, \frac{\partial F}{\partial v} \right)$. For~(\ref{EE1}) and~(\ref{EE2}) to hold identically (i.e. independent of how the branch is parametrized), the $v', u'$–dependence must cancel on the branch. This means that 
\[
A_1\left(-\frac{\partial F}{\partial u}\right)+B_1 \frac{\partial F}{\partial v} =0\qquad A_2\left(-\frac{\partial F}{\partial u}\right)+B_2 \frac{\partial F}{\partial v}=0
\]
which implies the relationship $ \displaystyle{ \frac{B_1}{A_1}=\frac{B_2}{A_2}}$ on the branch.  Cancelling the common factor $r\neq 0$, we then obtain 
\[
\frac{B_1}{A_1}=\frac{2v}{\,3v+4u\,}\qquad
\frac{B_2}{A_2}=\frac{v(v+2u+2)}{\,v^{2}+2vu+2v+2u\,}
\]
Equating these and simplifying yields the factorization
\[
\ v(v+2u)\,(v+4u+2)=0\ .
\]
Thus the only options for algebraic branches are
\[
v\equiv 0,\qquad v+2u\equiv 0,\qquad v+4u+2\equiv 0.
\]
A straightforward substitution shows that $v = 0$ and $v+2u = 0$  results in $C_1$ and $C_2$ vanishing identically. However the branch $v+4u + 2 = 0$ does not annihilate $C_1$ and $C_2$. Hence we conclude that only the branches $v = \mu' = 0$ and $v + 2u = \mu' + 2\nu' = 0$ are valid branches for the vacuum metric. There are no other possibilities. It is also easy to verify that these two branches satisfy the equation of pressure isotropy~(\ref{107d}). 

Further we can easily show that the Ricci tensor components vanish in both of these cases, however setting the Ricci tensor to vanish to find the vacuum metric as is done in general relativity would not have exposed the third option we had to consider. For this reason we said that the vanishing Ricci tensor would not be a necessary and sufficient requirement for the vacuum metrics. However, it turned out that the third option above proved to be defective thus leaving precisely two vacuum branches which we would have obtained from $R_{ab} = 0$ anyway. The approach we have followed here is rigorous and eliminates any other possibilities. Consequently~(\ref{vac1}) and~(\ref{vac2}) are the only possible vacuum solutions in pure quadratic $f(G, B)$ theory.

It is indeed remarkable that there are two vacuum solutions possible when in standard general relativity it is usual to have a single unique vacuum solution. For example, by Birkhoff's theorem the Schwarzschild exterior solution is the unique necessary and sufficient solution for the vacuum of a spherically symmetric fluid irrespective of whether the spacetime is static or not~\cite{Birkhoff,Weinberg}.

We now study each vacuum branch separately. 

\begin{itemize}
\item{(i)} If $\nu$ is a constant, meaning that the spatial slices are flat, then by~(\ref{107d}) we get $\mu'' =  - \frac{1}{2} \mu'^2$ which is solved by 
\be
e^{\mu} = B(r+A)^2 
\ee
so the metric applicable in this case is given by 
\be
ds^2 = -  Hdt^2 + B(r+A)^2(dx^2 + dy^2 + dz^2)   
\label{vac1}
\ee
where the integration constants have been redefined to $H$. The Kretschmann scalar for this metric vanishes. Given that the Ricci tensor is zero, in unusual coordinates, we can conclude that the conformal tensor also vanishes hence the metric is locally flat or Minkowskian. 

\item{(ii)} With $\mu = -\frac{1}{2}\nu + C_1$ equation~(\ref{107d}) reduces to the separable differential equation (essentially first order)  $\nu'' = - \frac{1}{4} \nu'^2$ with solution 
\be
e^{\nu} = C(r + A)^4 
\ee
and consequently
\be
e^{\mu} = E(r + A)^{-2} 
\ee
where $A, B, C, E$ are all integration constants. The associated metric has the form
\be
ds^2 = -C(r + A)^4   dt^2 +   E(r + A)^{-2} (dx^2 + dy^2 + dz^2). \label{vac2}
\ee
\end{itemize}
This metric has spatial slices that are conformally flat although the spacetime is curved in general. There is a singularity in the manifold at $r = -A$ and it will be interesting to classify this singularity. The Kretschmann invariant $K$ evaluates to 
\be
K = R_{abcd} R^{abcd} = \frac{192}{B^2(r+A)^{12}}. 
\ee
We can conclude that there is a curvature singularity of power-law type  at $r = -A$ and the spacetime is asymptotically flat as $r \rightarrow \infty$. This type of singularity, unlike a coordinate singularity which can be eliminated through gauge choices, is not removable. Note that a singularity in the vacuum spacetime is not unexpected - the well known Schwarzschild exterior solution of Einstein gravity is also singular at the centre of the distribution and another surface known as the horizon. 

We now critically assess the physical acceptability of each vacuum branch. For branch~(i), the Kretschmann scalar $K = R_{abcd}R^{abcd}$ vanishes identically, confirming that the spacetime is locally flat. The metric~(47) is simply Minkowski space written in non-standard isotropic coordinates, and it represents the natural flat vacuum background of the theory. This is entirely acceptable and plays the role analogous to the trivial Minkowski exterior in Newtonian gravity.
 
For branch~(ii), the Kretschmann scalar~(51) diverges as $(r+A)^{-12}$ at $r = -A$, demonstrating an irremovable curvature singularity at that location. This singularity is of power-law type and cannot be eliminated by any coordinate transformation; it is therefore a genuine spacetime pathology. A key difference from the Schwarzschild solution of general relativity is the absence of a Killing horizon: the lapse function $g_{tt} = -C(r+A)^4$ \emph{diverges} rather than vanishing at $r = -A$, so the singularity is not hidden behind a horizon. If $A < 0$ so that $r = -A > 0$ lies within the physical coordinate range, the singularity is naked in the sense of the Penrose cosmic censorship conjecture.
 
However, the physical significance of this singularity must be assessed in the context of its role as an \emph{exterior} vacuum metric. A realistic stellar model consists of a fluid interior matched to a vacuum exterior at a surface $r = r_b$ where the pressure vanishes. The junction conditions require continuity of the induced metric and of the extrinsic curvature at $r = r_b$. Provided the stellar radius satisfies $r_b > -A$ (which can always be arranged by an appropriate choice of the integration constant $A$), the singular point lies entirely outside the physical domain of the solution and plays no role in the stellar structure. The situation is precisely analogous to the Schwarzschild exterior, where the singularity at $r = 0$ is excised by the stellar interior; the difference here is merely that the singularity of branch~(ii) occurs at $r = -A$ rather than at the coordinate origin. The occurrence of richer vacuum structure --- two branches instead of the unique Schwarzschild solution guaranteed by Birkhoff's theorem in general relativity --- is a direct consequence of the higher-order nature of the $f(\mathcal{G})$ theory and underscores the genuinely new gravitational phenomenology accessible within this framework.

\section{Dynamical systems approach} 

We now revert to equation~(\ref{107d}) introduced in Section~\ref{sec3}. It may be observed that this is homogeneous under the transformations 
\[
r\mapsto \lambda r,\qquad u\mapsto \lambda^{-1}u,\qquad v\mapsto \lambda^{-1}v.
\]
for a real number $\lambda \neq 0$. In light of this, it is convenient to introduce the scale-invariant change of variables
\[
U(X) = r\,u(r),\qquad V(X)= r\,v(r),\qquad X =\ln r,
\]
so that $\mu' = u=U/r$ and $\nu' = v= V/r$. Then
\[
u'(r)=\frac{\dot U-U}{r^2},\qquad v'(r)=\frac{\dot V-V}{r^2},
\]
where $\dot{(\,)} =\frac{d}{dX}$. Substituting into~(\ref{107d})  and multiplying by $r^3$ yields an autonomous equation
\begin{multline}
    0 = 4(\dot U-U)\,V^2-2U^3(5V+1)-3U^4 + 
    U^2\!\Big(6\big[(\dot U-U)+(\dot V-V)\big]-7V^2-6V\Big) \\
    +2UV\Big(6(\dot U-U)+4(\dot V-V)+V^2-2V\Big).  
    \label{328}
\end{multline}
and there is no $r$ explicit dependence. This is remarkable and represents significant progress in understanding the underlying behaviour of the geometry of the spacetime. In standard general relativity it is not usually possible to reduce an isotropy equation to the autonomous form when studying astrophysical objects. In cosmological applications the dynamical systems approach is an often used method due to the number of equations and variables~\cite{Copeland1998}. In our context the reduction to an autonomous equation is helpful because if we let $\dot U=\dot V=0$ in~(\ref{328})  we can find the fixed points and show that they correspond to a scale-invariant profile
\[
u(r) \sim \frac{1}{r},\qquad v(r) \sim \frac{1}{r},
\]
which is very useful in finding exact solutions. We make it clear that the fixed points for this system are unique. The achievement of the forms for $u$ and $v$ are also important wins  for our analysis. 

Since we are dealing with a single equation containing two dependent variables, we can only progress by introducing additional assumptions. For example   assuming a proportionality between the metric variables on the invariant submanifold  $V=a\,U$ (so $\dot V=a\,\dot U$),~(\ref{328})
reduces to a first-order ODE for $U(X)$:
\[
6\dot U-8U+\frac{a^2-4a-3}{a+1}\,U^2=0\qquad (a\neq -1),
\]
which is  a logistic equation
\[
\;\dot U=\frac{4}{3}U-\frac{A}{6}\,U^2,\quad \;
\]
where $A=\frac{a^2-4a-3}{a+1}$ and whose general solution is straightforward to find. The logistic equation is a Ricatti equation that is easily solvable.  Transforming back gives
\[
u(r)=\frac{r^{1/3}}{\,C+\dfrac{A}{8}\,r^{4/3}},\qquad v(r)=a\,u(r),
\]
 a result we have obtained through the interpretation of the  logistic form. We will return to this exact solution later for a detailed study. 

Our objective in this work is to use the dynamical systems approach.  As we mentioned, we have a single equation. However it is possible to split the equation in two parts so we have a system of equations that we may implement the dynamical systems approach. This is desirable since our equation is autonomous. However the caveat is that there is no unique way to do the splitting.  Despite this there is immense value in performing such a split since the fixed points are unique and we will be able to infer the stability properties of the fixed points through the split. The underlying constraint equation~(\ref{328}) will still be obeyed. The dynamical systems approach is in any case used to understand the behaviour of solution trajectories near fixed points. It is not intended to locate exact solutions although in our case we have a solid suggestion of similarity structure (proportionality) that we can pursue for exact models. 

We now discuss a possible split. If we wish to treat~(\ref{328}) as a constraint relating $(U,V,\dot U,\dot V)$ we may write~(\ref{328}) as an autonomous system. The new coordinates we introduced had the distinct advantage of allowing us to write~(\ref{328}) in the form
\be
\alpha (U, V) \dot{U} + \beta (U, V) \dot{V} = - \gamma (U, V). \label{322}
\ee
This form allows us to make the interpretation from ~(\ref{322}) of $(\dot{U}, \dot{V})$ as a vector lying in the velocity plane. The set of all vectors that satisfy~(\ref{322}) is a line in the $(\dot{U}, \dot{V})$ plane since it has the form $\alpha \dot{U} + \beta \dot{V} = -\gamma$. A normal vector to that line would be $(\alpha, \beta)$ with direction any non-zero vector orthogonal to $(\alpha, \beta)$ such as $(-\beta, \alpha)$. We next need a gauge. In other words we need  a choice of one vector on that line. The common choice, amongst many options, is the minimal-norm or orthogonal projection of the form 
\be
\dot{U} = -\frac{\gamma \alpha}{\alpha^2 + \beta^2} \hspace{0.5cm} {\mbox{and}} \hspace{0.5cm} \dot{V} = -\frac{\gamma \beta}{\alpha^2 + \beta^2}
\ee
resulting in the shortest vector on the line that minimises $\dot{U}^2 + \dot{V}^2$ subject to the equation~(\ref{322}). Additionally this choice of gauge is ideal as it is smooth, bounded and the form of~(\ref{322}) allows us to read off fixed points when $\gamma = 0$. 

Before proceeding we address the question of gauge dependence. Any splitting of equation~(52) takes the general form $\alpha(U,V)\dot{U} + \beta(U,V)\dot{V} = -\gamma(U,V)$, and different choices of $(\alpha,\beta)$ produce different autonomous systems and hence different streamline portraits. However, three features of the analysis are \emph{gauge-invariant}. First, the fixed-point locus is determined entirely by the condition $\gamma(U,V) = 0$ (together with the overall prefactors of $\dot{U}$ and $\dot{V}$), which is independent of the choice of $(\alpha,\beta)$; consequently the fixed curves $U = 0$, $U+2V = 0$ and $Q(U,V) = 0$ are intrinsic to the autonomous equation~(52) and do not depend on the gauge. Second, the transverse stability eigenvalue $\lambda_\perp$ at a smooth point of a fixed curve can be expressed as $\lambda_\perp = -(\nabla\gamma \cdot \hat{n}) / |(\alpha,\beta) \cdot \hat{n}|$, where $\hat{n}$ is the unit normal to the fixed curve; for the minimal-norm gauge one has $(\alpha,\beta) \propto \hat{n}$, so $\lambda_\perp = -|\nabla\gamma|/|(\alpha,\beta)|$ which depends only on $\gamma$ and its gradient. Third, the sign of $\lambda_\perp$ --- and hence the classification of each segment as attracting or repelling --- is gauge-invariant, since a sign change would require the normal component of $(\alpha,\beta)$ to vanish, which is excluded away from isolated points. To confirm this in practice we have verified the stability classification using a second splitting, namely $\dot{U} = -\gamma/\alpha$ with $\dot{V} = 0$ (setting $\beta = 0$ locally away from $\alpha = 0$), and find identical stable and unstable segments on all three fixed curves. Thus while the visual appearance of the streamlines depends on the gauge, the qualitative conclusions regarding the stability properties of the invariant submanifolds are robust.

Applying this gauge we obtain the autonomous system 
\beq
    \dot{U} &=&  \frac{U(U+2V)(3U^2 +4UV+8U-V^2+8V)(3U^2 + 6UV + 2V^2)}{2\left((3U^2 + 6UV + 2V^2)^2 + U^2(3U + 4V)^2 \right)}  
    \label{331a} \\[1ex]
    \dot{V} &=& \frac{U^2(U+2V) (3U^2 +4UV+8U-V^2+8V)(3U+4V)}{2\left((3U^2 + 6UV + 2V^2)^2 + U^2(3U + 4V)^2 \right)} 
\label{331b}
\eeq
which will enable us to carry out a detailed phase plane analysis. Assuming the denominators remain non-zero, it can be seen that the fixed points, which they are if $U=V \neq 0$, are located on the invariant submanifolds $U = 0$, $U + 2V=0$ and $Q(U, V) = 3U^2 +4UV+8U-V^2+8V = 0$. The other factors $ U^2 + 6UV + 2V^2=0$ and $ 3U + 4V = 0$ do not lead to new invariant submanifolds but they add fixed points when they intersect one of the three submanifolds above. 

Let us now proceed to locate the fixed points by solving simultaneously:

\begin{itemize}
    \item{} Put $U = 0$ in $Q$ to give $-V^2 + 8V=0$ thus $V = \left\{0, 8\right\}$. So $(0, 0) $ and $(0, 8)$ are fixed points. 
    \item{} Inserting $V = -\frac{U}{2}$ in $Q = 0$ gives $3U^2 + 16U = 0$ hence $U = \left\{0, -\frac{16}{3}\right\}$. So we get an additional fixed point $\left(-\frac{16}{3}, \frac{8}{3} \right)$. 
\end{itemize}
These are not the only fixed points, however, we must bear in mind that the system is at rest along the entire invariant submanifold. These three points are the intersections of some of the branches. If we were to evaluate the Jacobian or stability matrix along the invariant submanifold, we would expect to generally find a zero eigenvalue. This is because the system remains at rest along the direction of the submanifold which will be reflected by the zero eigenvalue.

\begin{figure}[!htb]
    \centering
    \includegraphics[width=0.8\linewidth]{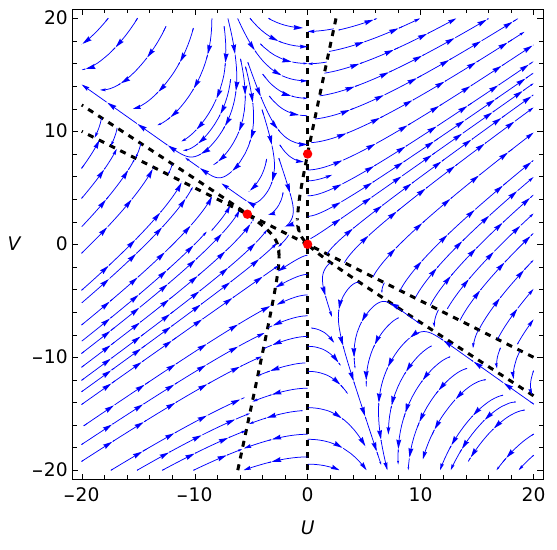}
    \caption{Phase space portrait of $\dot U,\dot V$ overlaid with invariant submanifolds $U=0$, $U+2V=0$, and $Q=0$ in dashed lines. The fixed point $(0,0)$, $(0,8)$, and $(-16/3,8/3)$ are in red.}
    \label{fig:5}
\end{figure}

As the system is non-linear we, find the Jacobian matrix $J$ to investigate the stability properties of the invariant submanifolds. We get 
\[
 J = 
\begin{bmatrix}
A & B \\
C & D 
\end{bmatrix}
\]
where
\beq
A &=& \left(81 U^9+27 U^8 (23 V+4)+144 U^7 V (14 V+5)+9 U^6 V^2 (403 V+224) \right. \n \\ && \left. +12 U^5 V^3 (331 V+260)+58 U^4 V^4 (47 V+52)+12 U^3 V^5 (95 V+168) \right. \n \\ && \left. 
+244 U^2 V^6 (V+4)+4 U V^7 (V+72)-4 (V-8) V^8\right) \n \\
&& / \left( 2 \left(9 U^4+30 U^3 V  
+32 U^2 V^2+12 U V^3+2 V^4 \right)^2  \right)   \label{334a} 
\eeq
\beq
B&=& \left(U \left(81 U^8+45 U^7 (11 V+4)+3 U^6 V (377 V+368)+24 U^5 V^2 (46 V+113)   \right. \right. \n  \\ && \left. \left. 
+10 U^4 V^3 (23 V+336)+2 U^3 V^4 (1068-175 V)+4 U^2 V^5 (152-59 V)
\right. \right. \n \\ && \left. \left. 
-48 U (V-1) V^6-4 V^8\right)\right) / \left( 2 \left(9 U^4+30 U^3 V+32 U^2 V^2+12 U V^3+2 V^4\right)^2\right) \n  \\  \label{334b} 
\eeq
\beq
C&=& \left(U \left(81 U^8+54 U^7 (11 V+2)+36 U^6 V (51 V+20)+6 U^5 V^2 (517 V+344)   \right. \right. \n \\ && \left. \left.
+2 U^4 V^3 (1543 V+1664)+4 U^3 V^4 (445 V+834)+4 U^2 V^5 (127 V+536)    \right. \right. \n \\ && \left. \left. 
+6 U V^6 (3 V+136)-16 (V-8) V^7\right)\right)  \n \\ &&   / \left(2 \left(9 U^4+30 U^3 V+32 U^2 V^2+12 U V^3+2 V^4\right)^2\right)  \label{334c} 
\eeq
\beq
D&=& \left(U^2 \left(54 (U+2) U^6+3 (87 U+176) U^5 V+2 (189 U+464) U^4 V^2  \right. \right. \n \\ && \left. \left.  -24 (U-24) U^3 V^3 
-4 (125 U+38) U^2 V^4-2 (35 U+32) V^6\right)\right)   \n \\ && -18 (21 U+16) U V^5
/ \left(2 \left(9 U^4+30 U^3 V+32 U^2 V^2+12 U V^3+2 V^4\right)^2\right) \n  \\ \label{334d}
\eeq
By direct substitution we find that 
\begin{itemize}
    \item{} On $U = 0$ the Jacobian becomes $ J = 
    \begin{bmatrix}
    4-\frac{V}{2} & 0 \\
    0 & 0 
    \end{bmatrix}$. This yields the eigenvalues $\lambda_1 = 0$ and $\lambda_2 = 4 - \frac{V}{2}$, any point on the submanifold. Therefore, the invariant submanifold $U = 0$ is unstable for $V < 8$ (this feature can be seen graphically if one focussed on this region of the phase portrait). When $V > 8$ the line attracts nearby trajectories, see Fig.~\ref{fig:5}.
    \item{} Along the invariant submanifold $U + 2V = 0$, the Jacobian has the form 
    \[  J = 
    \frac{1}{5}\begin{bmatrix}
    \frac{3U}{4} + 4 & \frac{3U}{2} +8 \\
    \frac{3U}{2} + 8 & 3U + 16 
    \end{bmatrix}\]
    Recall that one eigenvalue is expected to be zero and we find the other eigenvalue to be $\lambda_2 = \frac{3U}{4} + 4$. Hence, the submanifold is unstable for $U > -\frac{16}{3}$ and attracts nearby trajectories for $U < -\frac{16}{3}$, see Fig.~\ref{fig:5}.
    \item{} On the conic $3U^2 + 4UV + 8U - V^2 + 8V = 0$ the non-zero eigenvalue have the form \[\lambda_2 = \frac{A_1 F_1 + B_1 F_2}{A_1^2 + B_1^2} \] where $A_1 = 2(2V^2 + 6UV + 3U^2)$, $B_1 = 2U(3U + 4V)$ and $F_1 = \frac{\partial F}{\partial U} = -12U^3 - 30U^2V - 24U^2 -14UV^2 - 48UV + 2V^3 -16V^2$, $F_2 = \frac{\partial F}{\partial V} = 2U(-5U^2 - 7UV - 12U + 3V^2 - 16V)$ where $F = -U(U+2V)(3U^2 + 4UV + 8U - V^2 +8V)$ is the common parts of the numerators of $\dot{U}$ and $\dot{V}$. The sign of $\lambda_2$ determines the stability of the submanifold. It will be more instructive to see this depicted in a phase-plane portrait Fig.~\ref{fig:5}. 
\end{itemize}
Based on the above discussions, we obtain the phase-portrait shown in Fig.~\ref{fig:5}.

\subsection{Stability analysis based on phase diagram}

From our calculations,we have shown that at a non-degenerate equilibrium point on any invariant submanifold, the Jacobian of \((\dot U,\dot V)\) has rank one with eigenvalues \(\{0,\lambda_\perp\}\): the zero eigenvalue is tangent to the equilibrium curve (neutral motion along it), and the nonzero eigenvalue \(\lambda_\perp\) controls \emph{transverse} stability (returning toward or departing from the curve under small normal perturbations).

The sign of \(\lambda_\perp\) is \emph{gauge-invariant} and classifies the local normal behaviour. This suggests that 
\begin{description}
    \item{(i)} $
    \lambda_\perp<0 \;\Rightarrow\; \text{transversely attracting (stable)}$, 
    \item{(ii)}
    $\lambda_\perp>0 \;\Rightarrow\; \text{transversely repelling (unstable)}.
    $
\end{description}

From the phase space portrait Fig.~\ref{fig:5}, we can see that along \(U=0\) (the \(V\)-axis), $\lambda_{\perp} = 4 -\frac{V}{2}$ implies  
\begin{itemize}
\item stable for $V>8$,
\item unstable for $V<8$.
\end{itemize}

On the other hand along \(U+2V=0\) (i.e.\ \(V=-\tfrac12 U\)),
$\ \lambda_\perp=\tfrac{3U}{4}+4  $ implies 
\begin{itemize} 
\item stable for $U < -\frac{16}{3}$,
\item unstable for $U > -\frac{16}{3}$.
\end{itemize}

Additionally it may be observed that the conic \(Q(U,V)=0\) is a one-dimensional set of equilibria (a shifted hyperbola). Its transverse stability \(\lambda_\perp\) varies by segment; it changes sign precisely at the
intersection points \((0,8)\) and \(\big(-16/3,8/3\big)\) where \(\lambda_\perp=0\). In particular, in the third quadrant \((U<0,V<0)\) portions of \(Q=0\) are \emph{transversely attracting}, so trajectories are drawn toward the conic and then slide (neutrally) along it.

How do we interpret the behaviour of $U$ and $V$ with regards to $\mu$ and $\nu$ the original coordinates? Recall \(U=r\,\mu'(r)\), \(V=r\,\nu'(r)\), \(X=\ln r\). Transversely attracting segments of the fixed curves represent  similarity relations between \(\mu'\) and \(\nu'\) such that nearby trajectories return to these relations as \(r\) varies, while motion along the curve remains neutral that is: no point sink exists because equilibria are invariant submanifolds.

It is instructive to translate these phase-plane results back into the language of stellar physics. Recall that $U = r\mu'(r)$ and $V = r\nu'(r)$, where $\mu$ and $\nu$ are the isotropic-coordinate metric potentials appearing in the line element~(9). A trajectory that is attracted to the fixed curve $U = 0$ corresponds to $\mu'(r) \to 0$ as the independent variable $X = \ln r$ evolves, meaning that the spatial conformal factor $e^{\mu}$ asymptotes to a constant. From equation~(36) this implies $\kappa\rho \to 0$, consistent with an asymptotically flat exterior. Trajectories drawn toward the branch $U + 2V = 0$ (equivalently $\mu' = -2\nu'$) are those for which the temporal and spatial potentials are linked by the same relation that defines the second vacuum branch~(50); the energy density and pressure computed from~(36)--(37) along such trajectories are non-zero in general, but the spacetime approaches the vacuum geometry~(50) asymptotically. For the conic $Q(U,V) = 0$, transversely attracting segments in the third quadrant ($U < 0$, $V < 0$) correspond to geometries where both potentials are decreasing functions of $r$, which in turn implies via the Tolman--Oppenheimer--Volkoff-like equation~(23) a monotonically decreasing pressure profile --- the expected behaviour for a physically reasonable stellar interior. The compactness of the star, $\mathcal{C} = 2M/R$ where $M$ is the total gravitational mass and $R$ the stellar radius in isotropic coordinates, is determined once the boundary condition $p(r_b) = 0$ is imposed; trajectories that remain close to the attracting arcs of $Q = 0$ throughout the interior will generically produce moderate compactness values, although a quantitative bound requires the additional input of an equation of state. We refer the reader to~\cite{cruz} for a detailed treatment of the dynamical systems approach to relativistic stellar structure, including the physical interpretation of phase-space features.

\subsection{Poincar\'e compactification and fixed points at infinity}
 
To determine the fixed points at infinity we perform the standard Poincar\'e compactification. Let us introduce the variables
\begin{equation}
    X = \frac{U}{\sqrt{1+U^2+V^2}}, \qquad 
    Y = \frac{V}{\sqrt{1+U^2+V^2}}, \qquad 
    Z = \frac{1}{\sqrt{1+U^2+V^2}},
\end{equation}
which map the $(U,V)$-plane bijectively onto the open unit disk $0 \leq X^2+Y^2 < 1$. The boundary circle $X^2+Y^2 = 1$ represents infinity in the original variables. It will be convenient to parametrise points at infinity using the angle $\theta$ such that $X=\cos\theta$, $Y=\sin\theta$.
 
To identify the fixed points at infinity, we first note that the denominators of~\eqref{331a} and~\eqref{331b} are non-negative for $U=V \neq0 $ which means they only affect the scaling for the flow and hence can be ignored for global purposes. Next, we note that the right-hand sides of~\eqref{331a} and~\eqref{331b} are polynomials with highest degree of 6. Let us denote the relevant highest degree polynomials by $P_6$ and $Q_6$, respectively. The critical points at infinity are given by the roots of
\begin{equation}
    G_7 := X Q_6(X,Y) - Y P_6(X,Y) = 0,
\end{equation}
with $X^2+Y^2 = 1$. $G_7$, if not identically zero, will have at most 7 pairs of roots. Indeed, one finds 7 pairs of roots $\theta_i$ and $\theta_i + \pi$, for $i=1,\ldots,7$. We find
\begin{equation}
    G_7(\theta) = -\cos\theta(2\sin\theta+\cos\theta)
    \Bigl(2\sin(2\theta)+2\cos(2\theta )+1\Bigr)
    \Bigl(-5\sin\theta+3\sin(3\theta)+3\cos\theta+9\cos(3\theta)\Bigr),
\end{equation}
which determines the fixed points at infinity. 

\begin{figure}[!htb]
    \centering
    \includegraphics[width=0.8\linewidth]{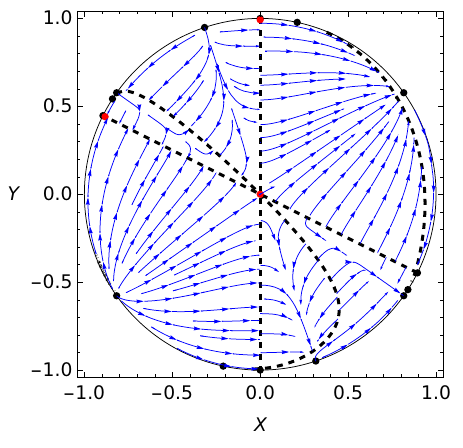}
    \caption{Global phase space portrait. The black dots denote critical points at infinity, red dots are the fixed points discussed in the local stability analysis. The dashed lines correspond to the invariant submanifolds.}
    \label{fig:global}
\end{figure}

The first quadrant of Fig.~\ref{fig:5} strongly suggests that there should be an attractor, possibly at infinity for large $U,V$. Likewise, we expect another attractor in the fourth quadrant for large  $U,-V$ and repellers in the second and third quadrants. The global phase space portrait Fig.~\ref{fig:global} largely confirms this, however, with significantly more details than the local analysis could ever reveal. There is a fixed point at infinity at $\theta_1 = \arctan(1/\sqrt{2})$, however, it is in fact the invariant submanifold (dashed line) which attract the trajectories. The point $\theta_1 + \pi$, on the other hand, acts as a repeller at infinity. All trajectories that emerge from this point terminate at an invariant submanifold. As one expects, the phase space trajectories change direction along the invariant submanifold, a feature that emerges beautifully in the global phase space portrait Fig.~\ref{fig:global}.

\section{Conclusion}

We have  studied the pure quadratic form of \( f(G, B) \) gravity as the next level of complexity since the linear order is precisely Einstein.  At the outset we examined the vacuum solutions associated with the theory by setting the energy-momentum tensor to vanish. It turned out that two solution trajectories with different geometric properties were admissible. One corresponds to a  spacetime with flat spatial slices and the other to a curved, asymptotically flat spacetime with an irremovable curvature singularity.  By rewriting the isotropy equation in a scale-invariant autonomous form, we were able to study the behaviour of solutions using phase diagrams. This was enabled by choosing a suitable gauge to split the master autonomous equation into two parts. Instead of isolated equilibrium points, the system admits curves of equilibrium states, which reflect the underlying symmetry of the problem. Although the split of the master equation is not unique, we gain considerable knowledge about the solution curves and their stability properties. The phase portraits show that some of these curves are stable in the sense that nearby solutions are naturally drawn toward them. This suggests  that the relationships between the metric functions are robust, making them good for exact and physically viable solutions. Overall, this approach gives a clear geometric picture of how the spacetime geometry behaves and  guides the construction of exact solutions. A further deduction possible from the dynamical systems approach is the suggestion that the metric potentials admit an inverse fall off in terms of the radius. A natural and important extension of this work is the construction of complete stellar models by supplementing the geometric framework developed here with a realistic equation of state. For instance, imposing a linear barotropic relation $p = \gamma \rho$ on the field equations~(36)--(37) generates an additional differential constraint on the metric potentials, selecting a one-parameter subfamily of trajectories from the phase portrait. Matching such interior solutions to the exterior vacuum metrics~(47) or~(50) at a boundary surface $r = r_b$ where $p(r_b) = 0$, together with the standard junction conditions on the induced metric and extrinsic curvature, will yield self-consistent models that can then be compared with the observed mass--radius relation of neutron stars. These investigations are in progress.

\end{document}